%% file: arxiv.tex
\PassOptionsToPackage{table}{xcolor}
\pdfoutput=1
\documentclass[sigconf,nonacm]{acmart}

\AtBeginDocument{%
  }

\setcopyright{acmlicensed}
\acmJournal{IMWUT}
\setcopyright{none}
\usepackage{subcaption}
\usepackage{makecell}
\usepackage{multirow}
\usepackage{multicol}
\usepackage{xspace}
\newcommand{\name}
{\textsc{GazeEval}\xspace}

\definecolor{bestcell}{HTML}{BBD6EC}
\definecolor{secondcell}{HTML}{E2EDF6}
\newcommand{\best}[1]{\cellcolor{bestcell}$\mathbf{#1}$}
\newcommand{\second}[1]{\cellcolor{secondcell}$#1$}

\usepackage{marvosym}
\makeatletter
\newcommand\correspondingauthor[1][Corresponding author]{%
  \if@ACM@anonymous\else
    \g@addto@macro\addresses{\g@addto@macro\@currentauthors{%
      \advance\hfuzz by 5pt\relax\textsuperscript{\Letter}}}%
    \g@addto@macro\@authornotes{%
      \begingroup\renewcommand\thefootnote{\Letter}\footnotetext{#1}\endgroup}%
  \fi}
\makeatother

\begin{document}

\title{Evaluating Visual Realism of AI-Generated Videos with \\Eye Tracking}

\author{Danqing Shi}
\email{danqing.shi@cs.lth.se}
\correspondingauthor
\affiliation{%
  \institution{Lund University}
  \country{Sweden}}

\author{Lan Jiang}
\email{lj450@cam.ac.uk}
\affiliation{%
  \institution{University of Cambridge}
  \country{United Kingdom}
  }

\author{Jake Lyons}
\email{jl2402@cam.ac.uk}
\affiliation{%
  \institution{University of Cambridge}
  \country{United Kingdom}}

\author{Miri Zilka}
\email{mz477@cam.ac.uk}
\affiliation{%
  \institution{University of Cambridge}
  \country{United Kingdom}}

\renewcommand{\shortauthors}{Shi et al.}

\input{Sections/0-Abstract}

\authorsaddresses{}
\maketitle
\pagestyle{plain}

\input{Sections/1-Introduction}
\input{Sections/2-Background}
\input{Sections/3-Study}
\input{Sections/4-Modeling}
\input{Sections/5-Conclusion}

\bibliographystyle{ACM-Reference-Format}
\bibliography{reference}

\input{Sections/appendix}

\end{document}

%% file: Sections/0-Abstract.tex
\begin{abstract}

Human evaluation of AI-generated videos has so far relied on explicit human feedback, such as judgments or ratings. While these signals capture the conclusions viewers reach, they fail to offer insights into the perception that leads to them.
This paper investigates eye tracking as a complementary human evaluation signal for AI-generated videos.
We conducted two Visual Turing Test (VTT) experiments with a total of 90 participants on videos generated by latest generative models, in which participants watched each video twice and then gave their feedback.
Across the experiments, we collected 4,286 valid eye-tracking trials on 438 videos.
Our analysis indicates that gaze patterns offer informative evaluation signals that align with explicit human feedback.
Building on the findings, we propose the first computational method \name to predict the perceived visual realism of AI-generated videos from eye-tracking data.
It achieves 71.3\% accuracy in prediction, providing evidence that eye tracking data carries informative signals about how people evaluate a video.
We release the study data to support further research on eye tracking for evaluating AI-generated video.
\end{abstract}

\ccsdesc[500]{Human-centered computing~Human computer interaction (HCI)}
\ccsdesc[300]{Human-centered computing~User studies}

\keywords{Human-AI Interaction, Eye Tracking, Video Generation, Human Perception, Generative AI}

\begin{teaserfigure}
  \centering
  \includegraphics[width=\textwidth]{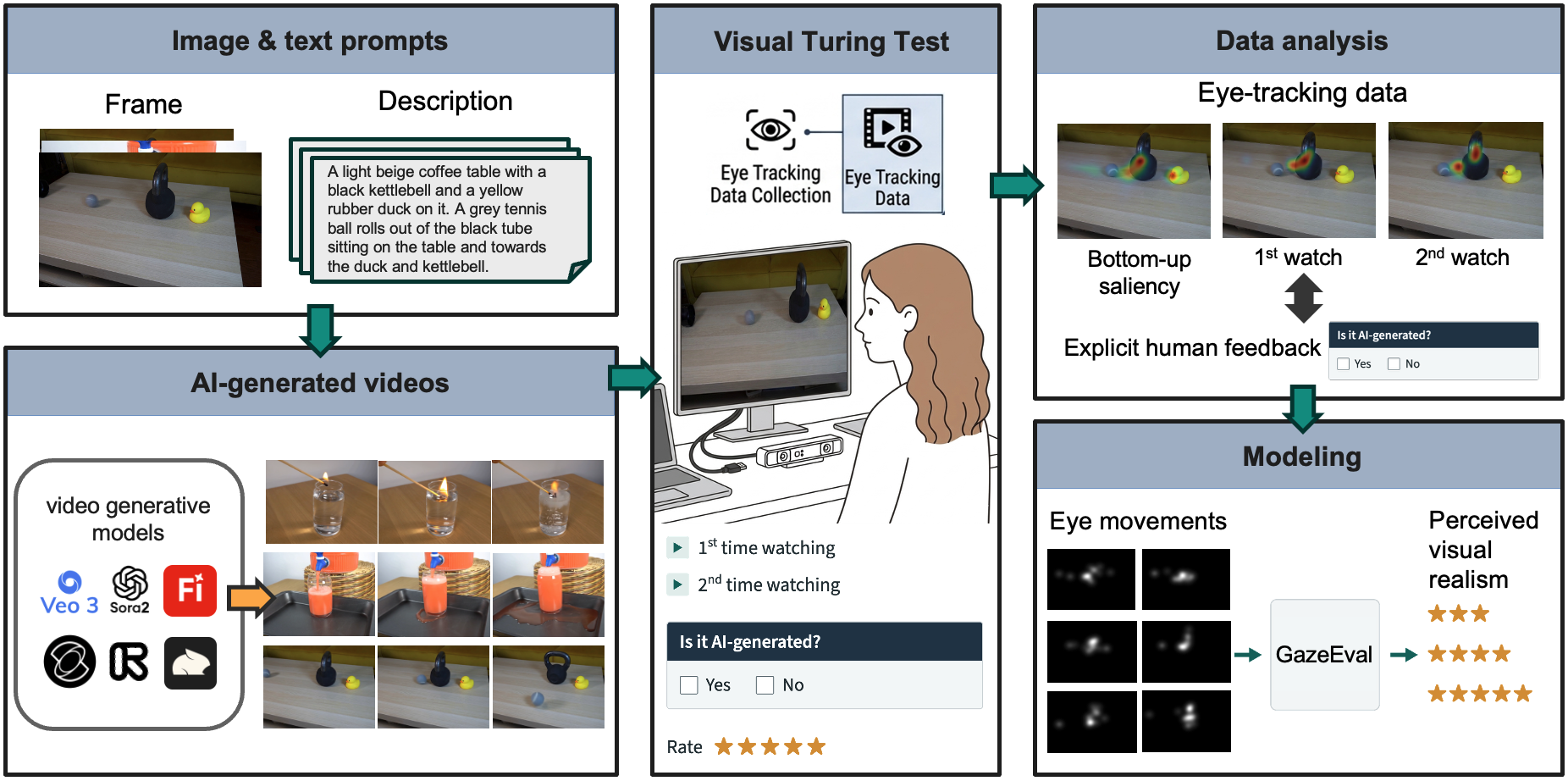}
  \caption{
  Our study investigates eye tracking as an evaluation signal for video generation. 
  \textit{Left}: The first frames and text descriptions of real videos are used for video generation with generative models, producing an AI-generated video set of physical scenes. 
  \textit{Middle}: Participants watch a mix of real and AI-generated videos while their eye movements are recorded under the Visual Turing Test (VTT): Participants are asked to judge whether each video is AI-generated and rate the visual realism after watching the video twice. 
  \textit{Right}: We analyze predicted saliency maps and eye-tracking data from the first and second watches. We train \name that predicts percieved visual realism from eye tracking data. Our findings indicate that eye tracking data from two-pass VTT can serve as an effective signal for video evaluation.}
  \Description{}
  \label{procedure}
\end{teaserfigure}

%% file: Sections/1-Introduction.tex
\section{Introduction}

Video generation has progressed rapidly in recent years, with generative video models such as Sora~\cite{sora} and Veo~\cite{google2025veo3} producing increasingly realistic content that is now widely shared and consumed on social media~\cite{renaultgrok, isthisAI}.
Evaluating these AI-generated videos becomes important, as it reveals the shortcomings of current models and guides their future development.
The evaluation of video generation can be divided into two categories: computational approaches and human evaluation approaches~\cite{huang2024vbench}.
Most computational approaches to evaluating video generation rely on automated metrics, including classical distributional measures such as FVD~\cite{unterthiner2019fvd}, alignment scores such as CLIP Similarity~\cite{radford2021clip}, and benchmark suites such as VBench~\cite{huang2024vbench} that aggregate many such indicators.
However, these metrics correlate only weakly with how people actually perceive video~\cite{otani2023toward, liu2023fetv}.
Human evaluation is therefore retained as the gold standard, and in practice it relies on explicit human feedback, such as binary ``real'' or ``fake'' decisions, Likert ratings, or pairwise preferences~\cite{zhang2024rethinking}.
Human visual perception itself is overlooked. In particular, human gaze behavior is a rich record of visual processing that can carry implicit signals of how a video is perceived before a viewer gives any explicit feedback.


In this paper, we investigate eye tracking as an implicit evaluation signal for AI-generated videos.
Gaze behavior analysis is a well-established research direction in cognitive science~\cite{hsiao2020role, tanaka2019utilizing}, and prior work has reported gaze differences between human-written and AI-generated text~\cite{ilyas2025reading, shi2024crtypist}, between real and edited static images~\cite{cartella2024unveiling, huang2025analysis}, and between real and face-swapped videos~\cite{gupta2020eyes, wohler2021towards}.
A recent study takes a first step toward eye tracking on AI-generated video, finding that authenticity shapes gaze behavior when people watch AI-generated videos of physical scenes~\cite{shi2026people, shi2026perceived}. However, the study was limited to analyzing low-level oculomotor statistics.
Furthermore, it remains open research questions whether gaze provides informative evaluation signals for video generation, and under which viewing conditions it can actually contribute to the evaluation of video generation.

To answer these questions, we first conducted a controlled eye-tracking user study with 40 participants from diverse backgrounds (\textit{batch 1}).
The experiment is designed as a Visual Turing Test (VTT) on videos, in which participants watch a video and then judge whether it is real or AI-generated.
The protocol lets participants watch each video twice, as we hypothesized that attention is first drawn to the saliency of the content and then shifts toward searching for anomalies during a second, verifying pass.
We analyze basic eye-tracking metrics, including the number of fixations, fixation duration, saccade magnitude, and scanpath length, across the different video conditions.
To account for the spatial movement of gaze, we further include eight saliency metrics~\cite{bylinskii2018different}, comparing the human attention map obtained in each pass against the bottom-up saliency predicted by a video saliency model~\cite{moradi2024salfom}.
Our findings reveal that human gaze behavior is affected by the visual realism of the videos.
Turning to the spatial movement of gaze, we find that gaze patterns are more consistent with bottom-up saliency in the first pass and shift toward more informative eye movements in the second.

We then propose \name --- the first computational method to predict perceived visual realism from human eye-tracking data.
To ensure generalization across video generative models, we collected a large-scale dataset by applying a similar VTT experiment to a broader video set of 300 videos generated by five mainstream video generative models, and recruited 50 participants (\textit{batch 2}) to watch these videos and provide explicit human feedback on them.
On this basis, we train models for predicting the perceived visual realism of AI-generated videos from eye-tracking data.
The best computational model achieves 71.3\% accuracy in prediction compared to the stimulus-only automated quality evaluation~\cite{huang2024vbench} at 61.4\%.
This result provides the evidence that gaze during two-pass watches in the experiment carries informative signals about how people evaluate a video.
Our ablation study shows that the saliency evaluation metrics are more informative than eye movement statistics, and the human self-consistency between the two watches is the key feature for the prediction of visual realism.

In summary, this paper explores eye tracking as an implicit signal for evaluating AI-generated video, beyond explicit human feedback.
Our main contributions are as follows:

\begin{itemize}
    \item We conduct Visual Turing Test experiments with eye tracking on videos generated by the latest video generative models with 90 human participants. Our analysis reveals that the visual realism of the videos significantly affects human gaze behavior under a two-pass watching protocol.
    \item We formalize evaluation of video generation with eye tracking as a prediction task, and present the first computational method \name for predicting visual realism from pure eye-tracking data. 
    It provides evidence that gaze during VTT carries informative signals for video evaluation. Our analysis identifies that human attention consistency across the two watches dominates the prediction of perceived visual realism.
    \item We release a publicly available dataset that contains 4,286 valid eye-tracking trials with accompanying explicit human feedback on the videos from advanced video generative models. To the best of our knowledge, no comparable dataset exists, and we intend it as a foundation for future research on perception-based evaluation of video generative models.
\end{itemize}

%% file: Sections/2-Background.tex
\section{Related Work}

This section reviews the literature on the evaluation of AI-generated videos, human evaluation protocols, and the intersection of generative AI and eye tracking.

\subsection{Evaluating AI-generated Videos}

Existing computational methods for evaluating AI-generated videos largely fall into automatic metrics~\cite{zhang2024rethinking}.
Automatic metrics dominate the literature: classical measures such as Inception Score~\cite{salimans2016improved}, Fr\'echet Inception Distance~\cite{heusel2017gans}, and Fr\'echet Video Distance~\cite{unterthiner2019fvd} compare distributions of generated and reference content, while CLIP Similarity~\cite{radford2021clip} estimates text--video alignment, and Video Quality Assessment models~\cite{tu2021ugc} predict perceptual quality from learned features.
Recent benchmark suites, including VBench~\cite{huang2024vbench}, EvalCrafter~\cite{liu2024evalcrafter}, and FETV~\cite{liu2023fetv}, aggregate multiple automated indicators across more fine-grained dimensions such as motion smoothness, temporal consistency, and prompt fidelity.
Despite their scalability, these metrics have well-documented limitations: they often depend on pre-trained backbones, require reference videos, are sensitive to surface-level statistics, and---most critically---correlate weakly with human perceptual judgments~\cite{otani2023toward, liu2023fetv}.
As a result, human evaluation remains the gold standard for evaluating the quality of AI-generated videos, which is the focus of this paper.

\subsection{Human Evaluation for Video Generation}

Recent benchmarks rely on human-rated data to calibrate or validate their automated scorers~\cite{huang2024vbench, liu2024evalcrafter}.
A recent survey of video generation~\cite{zhang2024rethinking} reveals substantial inconsistency in how human evaluation is conducted, raising concerns about reproducibility, reliability, and practicality.
Reported protocols vary widely across four key axes.
First, \emph{evaluation metrics} are inconsistently defined: some studies use a single ``overall preference'' score, while others decompose quality into more than ten dimensions; aspects such as temporal and motion quality have only recently received attention~\cite{huang2024vbench, liu2024evalcrafter}, and ethical considerations are rarely included~\cite{lee2023holistic}.
Second, \emph{evaluation methods} differ between absolute Likert-scale ratings and pairwise comparisons; while comparison-based protocols reduce annotator load, they typically rely on raw win rates that can be biased by uneven matchups and require many comparisons to yield stable rankings~\cite{rao1967ties}.
Third, \emph{annotators} are most often recruited directly by the authors rather than through crowdsourcing platforms, and many studies omit basic information such as the number of annotators, their training, or quality-control procedures~\cite{otani2023toward}.
Fourth, the choice and number of \emph{prompts} vary by orders of magnitude across studies, and small or unbalanced prompt sets can yield biased model rankings.
To address these issues, Zhang et al.~\cite{zhang2024rethinking} proposed the T2VHE protocol, which combines well-defined objective and subjective metrics, structured annotator training, pairwise comparison with a Rao--Kupper model~\cite{rao1967ties} for ranking, and a dynamic evaluation module that reduces annotation cost by roughly half.
Analogous efforts in text-to-image evaluation, such as HEIM~\cite{lee2023holistic} and the verifiable protocol of Otani et al.~\cite{otani2023toward}, similarly emphasize standardized instructions and quality control.
However, these protocols still rely on explicit judgments---ratings or preferences that annotators consciously report---and therefore inherit the well-known limitations of self-report: subjectivity, susceptibility to demand characteristics, and limited insight into how viewers actually perceive videos moment-by-moment.
Rather than refining the explicit human feedback pipeline, this paper investigates whether implicit human eye movements can serve as a human-centred signal for evaluating AI-generated videos.

\subsection{Generative AI Meets Eye Tracking}

Previous eye-tracking studies have investigated differences in cognitive processing using gaze data when comparing the reading of human-written and AI-generated text~\cite{ilyas2025reading},
proofreading behavior with and without AI assistant~\cite{shi2024crtypist},
human visual perception when individuals view partially forged images compared to authentic images~\cite{cartella2024unveiling},
and human perception of real and fake face images~\cite{huang2025analysis}.
Generally, these results in static content show that when people view AI content, their attention tends to focus on smaller and more targeted regions, as the AI content might be more predictable in terms of attention. 
Regarding videos, existing studies have primarily examined face-swapped videos~\cite{gupta2020eyes, wohler2021towards}.
The analysis of eye-tracking patterns showed differences in fixation behavior, with participants focusing more on the mouth area, which is usually unnatural during speech in face-swapped videos. 
However, the most common generation technique at that time was DeepFake~\cite{westerlund2019emergence}, which is much less advanced than recent generative AI models. 
Beyond authenticity, eye movements act as implicit signals indicating how viewers experience videos~\cite{zhu2022eyeqoe} and recognize emotions elicited by them~\cite{tabbaa2021vreed}. These studies demonstrate that eye movement data carries information about viewers' subjective experience of video content.
However, they focus on quality and emotional response rather than on the perceived visual realism of generated content.
The relationship between human gaze behavior and cognitive processes during engagement with real versus AI-generated videos remains insufficiently understood. 
To fill the gap, this paper studies eye tracking as a complementary evaluation signal for video generation.

%% file: Sections/3-Study.tex
\section{Visual Turing Test with Eye Tracking}
\label{sec:study1}

\subsection{Overview}

Existing human-evaluation protocols for video generation primarily depend on explicit human feedback from viewers, such as Likert ratings, pairwise preferences, or binary decisions regarding whether a video is real or fake~\cite{zhang2024rethinking}.
While the explicit human feedback is valuable, this data is sparse, and provides limited insights into the moment-by-moment visual processing that leads to it.
In this experiment, we pair explicit human feedback with implicit eye-tracking data collected while viewers watch the same videos.
This enables us to explore not only what conclusions viewers draw about a generated video, but also how people's watching behavior is shaped by the AI-generated video.

We adopt a Visual Turing Test (VTT)~\cite{GazeVaLM, 8363564}, which requires human participants to classify images as either real or AI-generated. In our experiment, we adapt this concept from images to videos. 
Different from prior study~\cite{shi2026people, shi2026perceived}, we use a two-pass protocol in which the viewer watches each video twice in sequence before they commit a judgement.
This repeated exposure to the same visual content is known to leave measurable traces in eye movements~\cite{bulling2011recognition}. 
The experiment design is designed to answer two research questions:

\begin{itemize}
    \item[\textbf{RQ1}] Does the visual realism of a video measurably change how people watch it?
    \item[\textbf{RQ2}] Do the watching behaviors in Visual Turing Test contribute valuable signals to tell the visual realism?
\end{itemize}

\subsection{Experiment Design}

\paragraph{Video stimuli}
The ground-truth real videos were drawn from the Physics-IQ dataset~\cite{motamed2025generative}. Each Physics-IQ video depicts a distinct physical scenario and is recorded at 30 frames per second.
The original dataset offers three videos for each scene captured from different angles. We randomly pick one of them for each scene as the original dataset. 
We remove all the static videos with no movement. It leads to 60 videos in total.
A comparable AI-generated video was prepared for each real-world video, so that every scene has a real video and an AI-generated version as a pair. 
To ensure similarity between real and AI-generated stimuli, we use the exact description of the scene in the original video dataset as the textual prompt, and the first frame of the original video as the image prompt. This ensures the AI-generated video has same physical phenomenon and share the same visual style. 
We used Google Veo~3.1~\cite{google2025veo3} as the generative model. All videos were cropped to remove visable AI-generation watermarks and trimmed to five seconds to match the length and motion of the real videos.

\paragraph{Participants}
We recruited $N{=}40$ adults (26 female, 18--70 years old, mean age 30.5) with normal or corrected-to-normal vision (\textit{batch 1} participants). Participants who wear glasses or contacts were allowed. One participant reported ADHD. Recruited participants were from interdisciplinary fields: social sciences and humanities (32.5\%), natural sciences (27.5\%), engineering (12.5\%), computer science (2.5\%), and other professional areas such as MBA, law, finance, and communications (25\%). The majority (75\%) had no prior experience using AI video-generation tools.
The university approved the ethics for the experiment, and informed consent was acquired from all participants.

\paragraph{Apparatus}
The experiment was conducted in a university lab.
We conducted experiments using the display-based eye-tracker, Gazepoint GP3 (60\,Hz). 
Stimuli were presented on a monitor with a resolution of $1680 \times 1050$ pixels. Participants were seated approximately around 60 cm from the monitor.
One instructor was seated nearby to assist with the experiment.
They were free to adjust their seat and the monitor before the experiment began. 
We used the Gazepoint Control software for calibration and the Gazepoint Analysis software to record gaze data.

\paragraph{Task}
\label{sec:tasks}
At the beginning of the experiment, participants were explicitly informed that some of the videos they will watch are real and some are AI-generated.
In each user session, the participant was assigned 20 videos, 10 of which are real and 10 are AI-generated.
Each video was played twice, with a short break in between.
After the end of each video, participants were asked to judge whether it was real or AI-generated. When they answered ``AI-generated'', they were asked what made the video look unrealistic to them.
Responses were given orally and transcribed by the experiment conductor.
Between trials and between video plays, a black screen with a central fixation cross was shown for three seconds to standardise the starting gaze position.
Session clips were assigned such that no participant saw both real or AI version of the same scene.

\subsection{Data Collection}

Gaze data were exported as fixations from GazePoint software, including the fixation positions and durations.
We setup the following rules to filter valid data:
A trial was retained only if (i) both passes of that trial were recorded, (ii) the recording spans at least four of the five seconds of the clip, and (iii) at least five fixations fall inside the stimulus.
The filtered two-pass data comprise 763 of the 800 possible (participant, clip) trials (95.4\%), each contributing exactly two viewings, for 1{,}526 eye-tracking trials in total.
Each eye tracking trial includes a median of about 12 valid fixations (min 5, max 18). 96.3\% of the recorded fixations survived the rule-based filters and enter our analysis.
The result data covers 78 clips (39 AI-generated, 39 real) drawn from 40 scenes, 38 of which are present in both variants.

\subsection{Eye Tracking Data Analysis}
\label{sec:metrics}

We divided the analysis into two aspects. 
Eye movement statistics summarize how the eyes moved, regardless of the content.
Attention maps summarize where gaze landed, and they can be evaluated by comparing one attention map to another using saliency metrics.

\paragraph{Eye movement statistics.}
The four measures~\cite{gupta2020eyes} below are computed per viewing:
\textit{Number of fixations} measures total fixations recorded during a trial. A higher count indicates more information sampling.
\textit{Fixation duration} measures average duration of fixations in seconds. Longer durations indicate more effortful processing at single locations~\cite{ilyas2025reading}.
\textit{Saccade magnitude} measures distance between consecutive fixations. Greater saccade magnitude reflects shifts in attention across more distant regions and can reflect the efficiency of visual processing~\cite{pannasch2008visual}.
Saccades are taken as the displacement between consecutive fixations in frame pixels, so that the first fixation of a viewing contributes no saccade---the movement that reached it was not made during this viewing.
\textit{Scanpath length} measures total distance traversed by the user's scanpath across the video. A longer scanpath indicates a more exploratory viewing pattern.
These four metrics are not independent and should be interpreted as a single description of viewing behavior rather than as separate tests. 
Scanpath length is nearly equal to the product of mean saccade magnitude and fixation count. Since a five-second clip has a fixed time limit, increasing the number of fixations results in shorter fixation durations.

\paragraph{Human attention maps.}
To identify where the gaze landed, we compute a single attention map from all five seconds of each eye tracking trial.
Only fixations within the stimulus area are used to build the attention map, and those falling outside the viewing window are considered tracking loss and therefore excluded.
Each fixation is weighted based on its duration, so a fixation held for 400\,ms contributes twice what one held for 200\,ms does.
The resulting map is then convolved with an isotropic Gaussian of $\sigma = 0.03$ of the frame width, which corresponds to roughly one degree of visual angle at our viewing distance.

\paragraph{Bottom-up saliency prediction.}
For our reference on stimulus-driven attention, we utilize SalFoM~\cite{moradi2024salfom}, a video saliency foundation model that predicts a saliency map for each frame of the video independently. 
The per-frame predictions are combined over the entire clip onto a consistent $960 \times 720$ grid, resulting in one predicted map per clip that aligns with the human attention maps.
As a result, SalFoM represents ``where an average viewer would look, given only what the clip looks like'' without any specific tasks involved.

\paragraph{Saliency evaluation metrics.}
We score one attention map against another to measure the alignment using the eight standard saliency benchmark metrics~\cite{bylinskii2018different}, computed once per viewing on the whole-clip maps:
\begin{itemize}
    \item \textit{Area Under ROC Curve (AUC)}~\cite{fawcett2006introduction} evaluates saliency as a classifier of fixations: how well the map separates fixated from uniformly sampled locations (higher is more aligned).
    \item \textit{Shuffled AUC (sAUC)}~\cite{borji2013analysis} as AUC, but with negatives drawn from the fixations of other clips, which discounts the centre bias (higher is more aligned).
    \item \textit{Normalized Scanpath Saliency (NSS)}~\cite{peters2005components} measures the saliency at fixations, computed as the average normalized saliency at fixated locations (higher is more aligned).
    \item \textit{Pearson’s Correlation Coefficient (CC)}~\cite{le2007predicting} evaluates the linear relationship between distributions, measuring how correlated or dependent two variables are (higher is more aligned).
    \item \textit{Earth Mover’s Distance  (EMD)}~\cite{rubner2000earth} incorporates spatial distance into evaluation, computed by the cost of transporting one density onto the other (lower is more aligned).
    \item \textit{Similarity (SIM)}~\cite{rubner2000earth} measures the intersection between distributions, computed by histogram intersection of the two densities (higher is more aligned).
    \item \textit{Kullback--Leibler Divergence (KL)}~\cite{emami2013selection} evaluates saliency with a probabilistic interpretation (lower is more aligned).
    \item \textit{Information Gain (IG)}~\cite{kummerer2015information} evaluates information gain over a centre-bias baseline (higher is more aligned).
\end{itemize}
We analyze these metrics through three comparisons, as illustrated in Figure~\ref{full_analysis}. 
\textbf{(a) 1st vs.\ 2nd watch} evaluates how consistently an individual focused on the same clip during two separate viewings.
\textbf{(b) 1st watch vs.\ SalFoM} and \textbf{(c) 2nd watch vs.\ SalFoM} evaluates the model's predictions against the viewer's attention patterns during each viewing, asking how ordinary a viewer's attention was given the content alone.

\subsection{Findings}

\begin{figure*}[!t]
\centering
\includegraphics[width=0.7\linewidth]{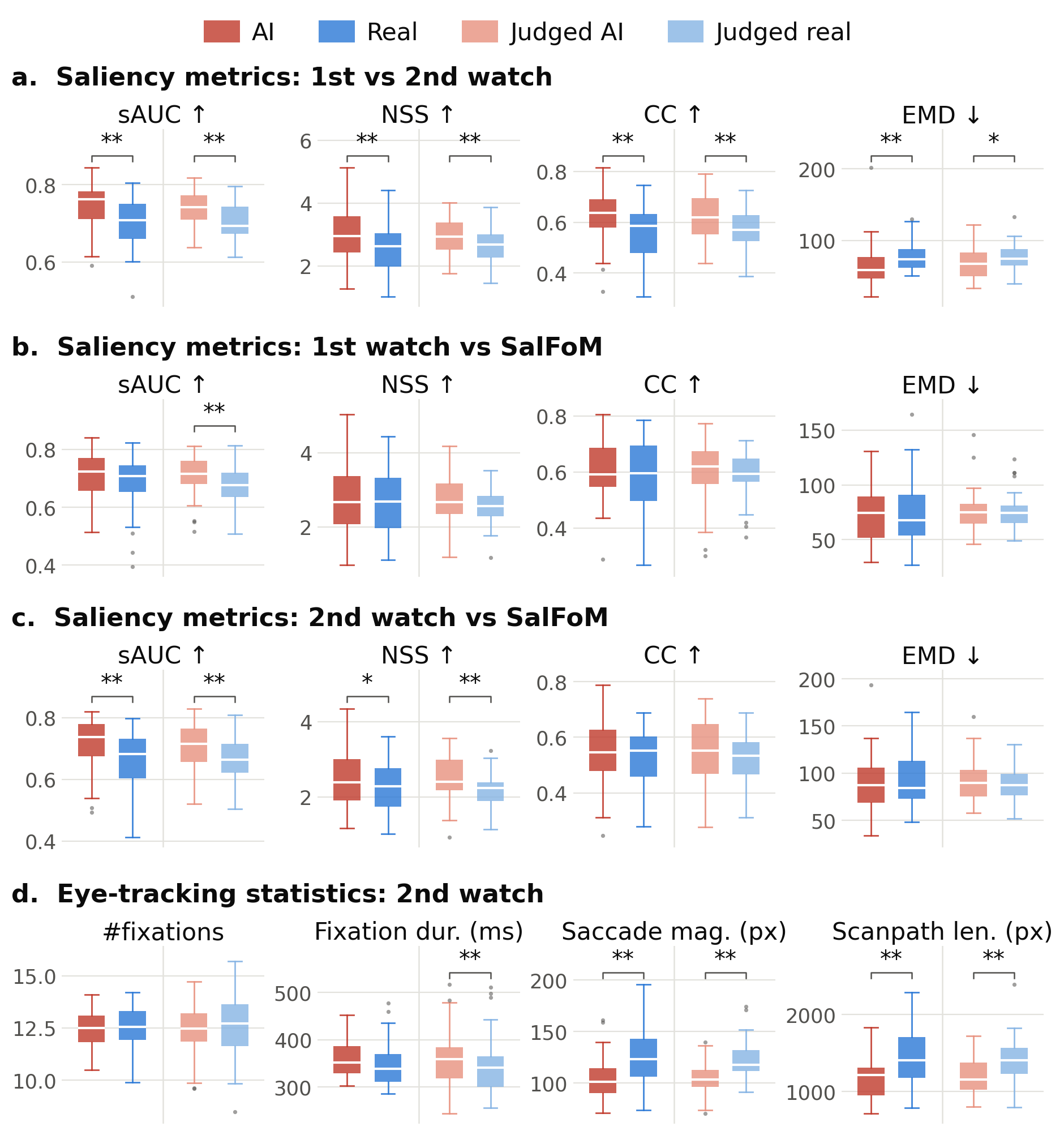}
\caption{Group comparisons for the two-pass Visual Turing Test. (a--c) report the four saliency evaluation metrics for the three attention-map comparisons; (d) report the four eye movement statistics for the second watch. 
Each panel carries two comparisons: AI vs.\ real clips and judged AI vs.\ judged real. Brackets mark Wilcoxon signed-rank tests: $^{*}\,p<.05$, $^{**}\,p<.01$. The full report can be found in Appendix~\ref{sec:appendix_stats}.}
\label{full_analysis}
\end{figure*}

Overall, the accuracy of human judgements in the the two-pass Visual Turing Test achieves 78.4\%.
The results of group comparisons are illustrated in Figure~\ref{full_analysis}, focusing on both ground-truth realism and perceived visual realism. The main findings are as follows:

\paragraph{AI-generated clips elicit markedly more self-consistent gaze across the two passes.}
The comparison of viewers' two-pass watches (see Figure~\ref{full_analysis}-a) reveals a clear distinction between the two conditions: all saliency metrics demonstrate significant differences in the same direction ($p<.05$). 
This indicates that viewers who watch either an AI-generated clip or a judged AI clip twice tend to look at the same areas both times, while viewers watching a real clip do not exhibit this consistency. 
One explanation is that AI-generated clips offer less to come back to: the second look finds the same few salient regions, whereas a real scene continues to reward exploration. 
This is the most significant consistent effect observed in this experiment. This finding can only be captured by the watch protocol in our two-pass Visual Turing Test experiment design.

\begin{figure*}[!t]
\centering
\includegraphics[width=0.8\linewidth]{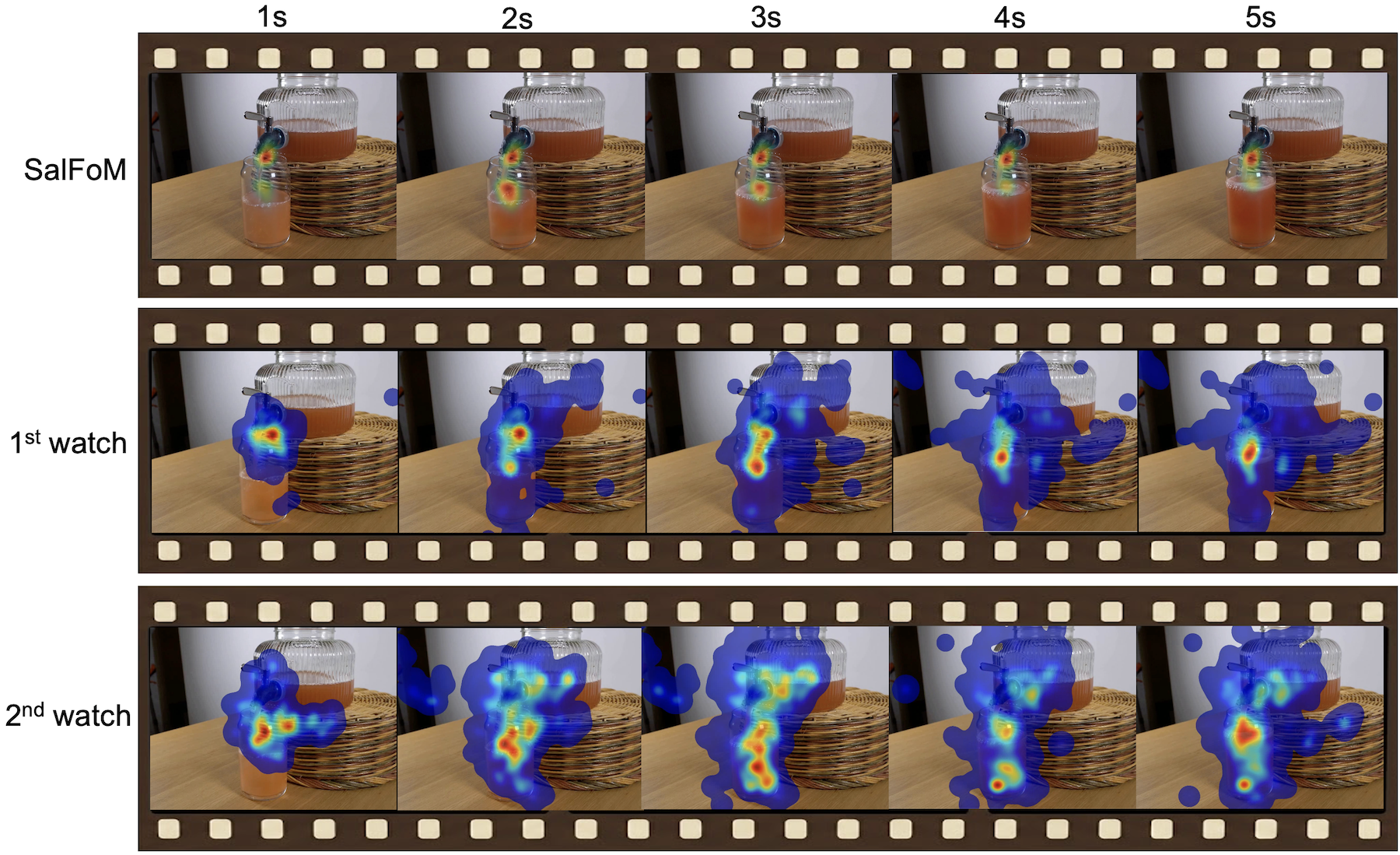}
\caption{An illustration of five frames (one per second) from a video generated by the prompt ``\textit{A glass beverage dispenser pouring grapefruit juice into a glass that contains some water.}''  The clip is decomposed second by second here to make the temporal pattern visible. 
The top row shows the bottom-up dynamic video saliency predicted by SalFoM, focusing on the motion area.
The other two rows shows the aggregated attention maps from the first and second watch. The human attention map during the first watch aligns more closely with the bottom-up saliency map, while the attention map from the second watch shows greater divergence, as viewers attempt to identify anomalies.
In this video, the error is that while the glass dispenser pours out water, the water level remains unchanged. During the second watch, viewers focused more on the water level.} 
\label{case}
\end{figure*}

\paragraph{The first pass follows bottom-up saliency, but the second deviates from it.}
Comparing the agreement in saliency metrics with SalFoM is generally lower in the second pass than in the first. This is evident in the SIM, CC, NSS, and AUC metrics for both stimulus types, with both distances increasing. Specifically, mean CC falls from $0.600$ to $0.547$ on AI clips and from $0.591$ to $0.525$ on real ones, and mean EMD rises from about $76$\,px in the first pass to $89$ and $91$\,px in the second. 
When a viewer watches a clip for the first time, their attention is largely guided by the clip's saliency. However, during the second watch, their gaze behaviors become less predictable, as they may actively seek out anomalies in the video due to AI-generated content. 
Figure~\ref{case} illustrates this pattern for a clip generated by the prompt: ``\textit{A glass beverage dispenser pouring grapefruit juice into a glass that contains some water.}''
On the first watch, viewers focus on the key area of motion where the stream of the liquid being poured, which aligns with the predictions made by SalFoM. 
In contrast, during the second watch, viewers shift their attention to the anomaly in the video: while the dispenser pours, the water level in the glass remains unchanged. As a result, viewers are more aware of the water level in the dispenser during their second watch.

\paragraph{Real videos are explored more widely, and AI-generated videos are dwelled upon.}
The eye movement statistics (see Figure~\ref{full_analysis} d) indicate a consistent pattern across both viewing sessions. Saccade magnitude is shorter for AI-generated clips during the first pass and is 17\% shorter during the second pass. Similarly, scanpath length mirrors this trend, showing reductions of 14\% and 19\%, respectively.
Viewers watch more of the frame when the clip is real or judged real, and this gap increases during the second watch.
On the other hand, fixation duration is about 20\,ms longer on AI-generated clips in the first pass ($361.8$ vs.\ $341.4$\,ms) and shows a similar but non-significant trend in the second pass ($362.5$ vs.\ $347.4$\,ms). The number of fixations does not change significantly in either viewing session.
Given the fixed five seconds, the more fixations and shorter saccades represent two perspectives of the same behavior, indicating that attention is focused on fewer regions. 
Notably, the effects of the differences between AI-generated and real videos are significant even during the first watch, prior to viewers having chance to verify the realism in the second watch.

\paragraph{Similar phenomena are observed in both ground-truth realism and perceived visual realism}
Grouping viewings based on participants' judgments rather than ground truth closely reproduces the original pattern (see the comparison between the dark and light box plots in Figure~\ref{full_analysis}).
Out of the 18 significant results in judgment-based comparisons, 15 have a corresponding significant result ($p<.05$) when categorized by stimulus. The remaining three results show similar trends, although they are not significant. 
This indicates that human evaluation reflect not only the actual content of the video but also their perceptions of the video. 
This aligns with the purpose of a perception-based evaluation signal, as human evaluation of generated video is a judgment about perceived realism rather than its ground-truth realism.

\paragraph{Summary.}
Answering ``\textit{Does the visual realism of a video measurably change how people watch it?}'' (RQ1): The visual realism of a video significantly affects how it is viewed. Real content encourages broader, more exploratory scanning, while AI-generated content captures attention quickly and tends to return to the same areas.
Answering ``\textit{Do the watching behaviors in Visual Turing Test contribute valuable signals to tell the visual realism?}'' (RQ2): The human attention maps gaze behaviors show valuable information in two-pass watches. The first watch is more closer to bottom-up saliency than the second, and their consistency with each other is a strong signal we can measure. 
This motivates the next section, which asks whether these signals are strong enough to predict visual realism of AI-generated video.

%% file: Sections/4-Modeling.tex
\section{\name: From Eye Tracking to Perceived Visual Realism}
\label{sec:study2}

Based on the findings from Section~\ref{sec:study1}, this section explores whether eye tracking data can serve as a useful signal for video evaluation. 
The central question addressed here is: can we predict how realistic an AI-generated video appeared to viewers?
To this end, we propose a computational method \name to predict visual realism from eye tracking data.

\subsection{Data Collection}

\paragraph{Video stimuli}
To make our findings generalizable beyond a single generator, we extended the stimulus set to five video generative models available on the Adobe Firefly platform for generative AI~\footnote{https://www.adobe.com/products/firefly}: Firefly, Sora~2, Pika~2.2, Runway Gen-4.5, and Kling~3.0.
We excluded Veo~3.1, which was used in the experiment of Section~\ref{sec:study1}, so that the modeling data and the analysis data share no generator.
The same Physics-IQ scenes were used as sources, and each was rendered by all five generators under the prompting protocol of Section~\ref{sec:study1}---the scene description as the text prompt and the first frame of the real video as the visual anchor. This led to 300 AI-generated video clips alongside the 60 real original clips, for 360 clips in total.

\paragraph{Participants and procedure}
We recruited $N{=}50$ additional adults (26 female, 24 male; Mage = 29.5, SD = 10.1, range 19–60) with normal or corrected-to-normal vision (\textit{batch 2} participants), none of whom took part in the experiment of Section~\ref{sec:study1}.
The apparatus, calibration and trial structure were similar to the two-pass VTT described in Section~\ref{sec:tasks}: each clip was played twice, after which the participant judged whether it was real or AI-generated and, when they answered ``AI-generated'', rated its visual realism on a five-point scale.
Each participant watched 60 clips sampled from the pool, so that every AI-generated clip was seen by about 8---9 viewers.

\paragraph{Data collection}

\begin{figure*}[!t]
\centering
\includegraphics[width=0.95\linewidth]{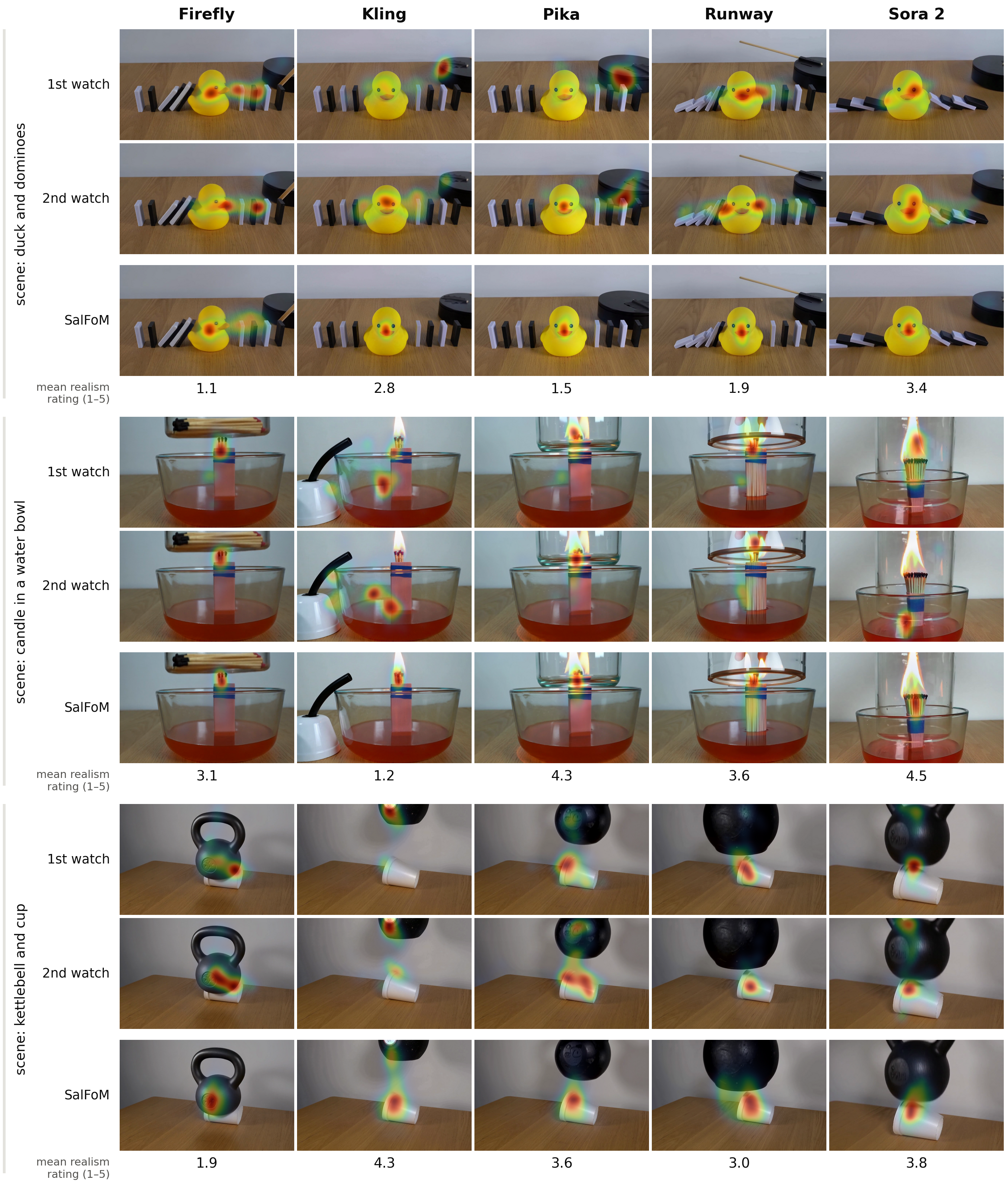}
\caption{Sample videos from five video generative models with human attention maps from our data collection. All the study videos and human attention maps can be found in the supplemental material.}
\label{dataset}
\end{figure*}

After applying the same filtering process described in Section~\ref{sec:study1}, we obtained a total of 2,760 valid eye-tracking trials.
Since we are focusing on AI-generated videos, we included 2,343 (viewer, clip) pairs from the 300 AI-generated clips for the modeling process.
To establish labels for perceived visual realism in our modeling, we combined binary judgments and graded ratings into a single five-point visual realism rating in the following manner:
a viewer who identified a clip as AI-generated provided their rating directly. In contrast, a viewer who was deceived into thinking the clip was real did not enter a rating, and we assigned them the maximum score, as being fooled is the strongest assertion that a clip appears realistic.
The perceived visual realism of a video clip is calculated as the average of these ratings from all viewers.
Figure~\ref{dataset} presents samples of videos from five video generative models with averaged attention maps in our data collection.

\subsection{Do Gaze Agreement Metrics Track Perceived Realism?}

\begin{figure}[!t]
\centering
\includegraphics[width=\linewidth]{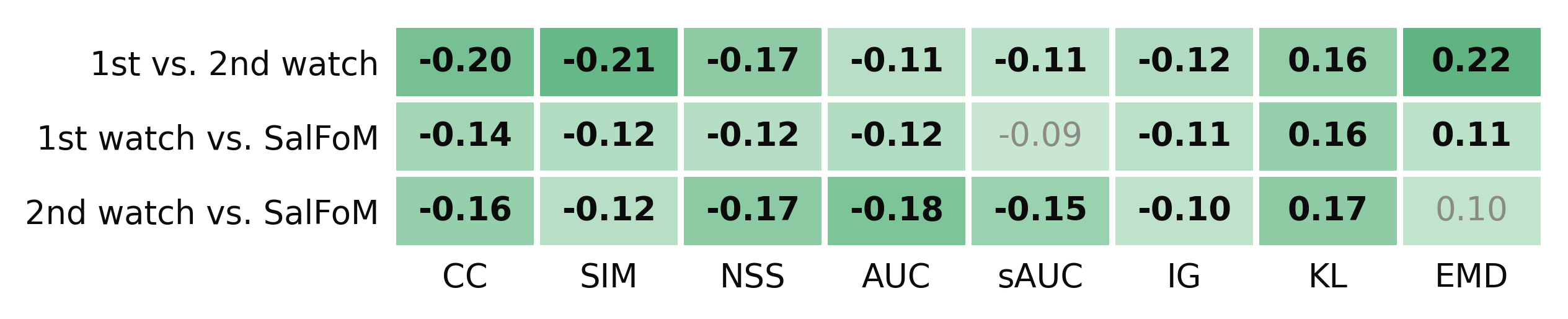}
\caption{The Spearman correlation was calculated between each saliency evaluation metric and viewers’ own five-point visual realism ratings across all (viewer, clip) pairs for the 300 AI-generated clips. The analysis includes three types of attention-map comparisons. Six metrics (CC, SIM, NSS, AUC, sAUC, IG) measure agreement, while two (KL and EMD) measure distance, leading to an expectation of opposite signs for their correlations. In all cases, a greater consistency in gaze patterns corresponds with videos that are perceived as less realistic.}
\label{fig:correlation_analysis}
\end{figure}

Before building a prediction model, we ask whether the quantities identified in Section~\ref{sec:study1} relate to the rating at all.
Figure~\ref{fig:correlation_analysis} reports the Spearman correlation between each saliency evaluation metric and the viewer's own realism response, over every (viewer, clip) pair, for the same three comparisons as before.
The six agreement metrics (CC, SIM, NSS, AUC, sAUC, IG) correlate negatively with perceived realism and the two distances (KL and EMD) correlate positively.
It means the more consistent a viewer's attention is, both with the bottom-up prediction, and with their own earlier pass, the less real the clip looked to them.
The strongest correlation is found in the human self-consistency row (SIM $\rho{=}-0.21$, CC $\rho{=}-0.20$, EMD $\rho{=}0.22$), and within the two SalFoM rows the second pass is consistently stronger than the first (for example AUC $-0.18$ vs.\ $-0.12$, NSS $-0.17$ vs.\ $-0.12$).
This order aligns with the findings in Section~\ref{sec:study1}: self-consistency was the comparison that most distinctly separated the conditions.
However, all the correlations themselves are modest, between 0.1 and 0.22.
Both the metrics and the ratings used in this correlation analysis are somewhat noisy, as the average number of fixations per video per viewer is only about 12, and individual preferences vary significantly. 
As a result, the modeling is conducted at the video level, where averaging the responses over multiple viewers helps to reduce much of this noise.

\subsection{Predicting Perceived Visual Realism from Eye Tracking Data}

\begin{figure*}[!tp]
\centering
\includegraphics[width=\linewidth]{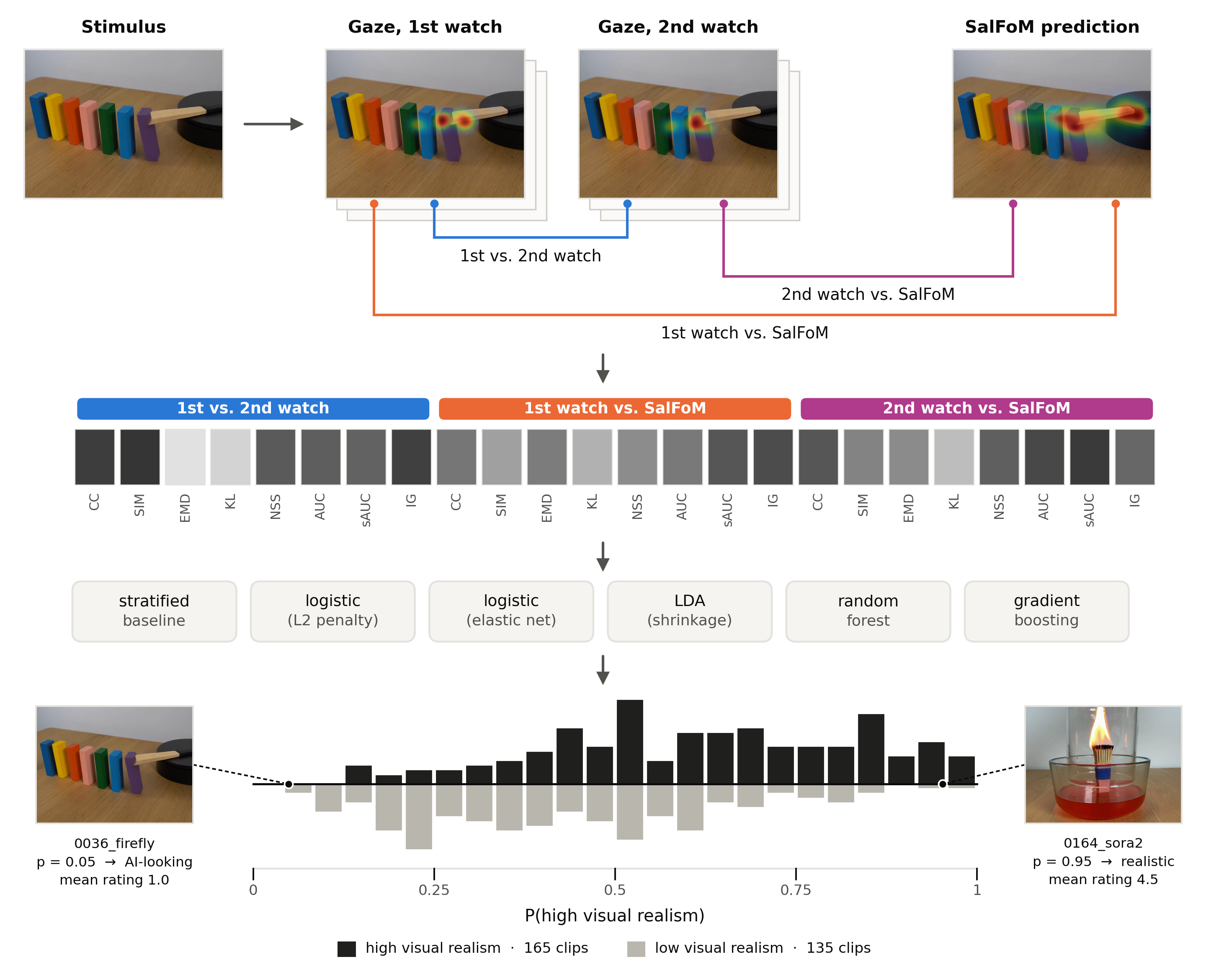}
\caption{\textbf{From eye tracking data to visual realism prediction.}
\emph{Top:} This presents a running example. Accompanying the stimulus keyframe are the attention maps from viewers' first and second watches, each generated by pooling the entire clip. On the right, SalFoM's prediction for the same clip is shown, derived from the raw pixels. The three comparisons form the basis for the model inputs: the first watch compared with the second watch, and each of the two watches compared against SalFoM's prediction.
\emph{Middle:} Each comparison is evaluated using eight saliency metrics: CC, SIM, EMD, KL, NSS, AUC, sAUC, and IG, resulting in the 24 video-level features. Since SalFoM generates a density map without attached fixations, the directed metrics are applied in one direction only. The resulting vector feeds into a classifier from the six candidate families listed, utilizing cross-validated ROC-AUC within each training fold.
\emph{Bottom:} This histogram represents the held-out probabilities for all 300 clips, categorized by the ratings the clips actually received. The bars above the line indicate the 165 clips with a mean rating exceeding 3, while the bars below indicate the 135 clips that did not. Each clip is scored based on the fold that held it out within a single scene-grouped partition.}
\label{fig:modeling}
\end{figure*}

Figure~\ref{fig:modeling} illustrates the pipeline of \name.
A video enters as a set of viewings. Each viewing is computed to an attention map. The maps are turned into agreement scores by the three comparisons of Section~\ref{sec:metrics}. The scores are averaged over the clip's viewers to give one feature vector per video, and the model are trained to map that vector to a probability that the video looked real.

\paragraph{Dataset}
The dataset used for modeling consists of a total of 300 data points, with each data point representing one AI-generated video and its associated viewing data from human participants.
To label the visual realism of the videos, we utilize an average rating threshold of 3.0. A video is considered realistic if its average viewer rating exceeds this midpoint. In the dataset, 165 of the 300 videos (55\%) received ratings above this threshold.
Each of the 60 source scenes was rendered by all five generators, meaning that two clips can have the same content. If a split is made on a clip, it may result in one scene being included in the training set under one generator and in the test set under another. This could allow the model to identify the scene itself rather than recognizing viewer gaze.
To ensure consistency, folds are grouped by source scene, ensuring that all five renderings of a particular scene remain on the same side of every data split. We use five folds, creating an 80/20 train/test division, and construct them so that each fold maintains the overall class ratio of approximately 55\%. 

\paragraph{Baselines}

We included two baseline methods in our analysis. 
The first is a trivial baseline that always predicts the majority class, achieving an accuracy of 0.550, as well as a balanced accuracy and ROC-AUC of 0.500.
The second baseline is a stimulus-only approach, which evaluates a video based solely on its own characteristics, using the dimensions outlined in VBench~\cite{huang2024vbench}. We compute seven video quality features: Subject Consistency, Background Consistency, Temporal Flickering, Motion Smoothness, Dynamic Degree, Aesthetic Quality, and Imaging Quality. The models are trained in the same manner as previously described.
We do not include measurement features that assess a video against its generation prompt, as viewers never saw the prompts and evaluated each clip based on the video content only.

\paragraph{Features}
The three comparisons at the top of Figure~\ref{fig:modeling} are the three blocks of the feature vector below them, each contributing the same eight saliency evaluation metrics (CC, SIM, EMD, KL, NSS, AUC, sAUC, IG):
\emph{1st vs.\ 2nd watch} scores a viewer's two passes against each other---how repeatable that individual's attention was.
\emph{1st watch vs.\ SalFoM} and \emph{2nd watch vs.\ SalFoM} score the stimulus-driven prediction against each pass---how ordinary that viewer's attention was, given only the content.
Each comparison is scored per viewer and then averaged over the clip's viewers. This is because pooling all viewers into a single attention map before measuring costs about 0.07 ROC-AUC, due to between-viewer disagreement that pooling integrates away is itself predictive of how real a video looked.
The union of the three blocks gives the 24 video-level features the figure shows, and this is the model reported as the best row of Table~\ref{tab:prediction}.
For the ablation we additionally fit each block on its own, an eye movement statistics block of eight features (fixation count, mean fixation duration, mean saccade magnitude and scanpath length, on each of the two passes, likewise averaged over viewers).
We also include the union of saliency features and eight statistic features ($n=32$), and the union of saliency features and seven VBench features from video stimulus ($n=31$).

\paragraph{Models}
In the model layer, we include six candidates: a stratified baseline classifier, $\ell_2$- and elastic-net-penalised logistic regression, shrinkage linear discriminant analysis (LDA), a random forest and gradient boosting. This selection includes the estimator classes allowed within this scope rather than exhaustively listing all possibilities. 
Model selection, hyperparameters, and the decision threshold are all determined using cross-validation within the training dataset.
The selection is scored based on ROC-AUC, which is threshold-free and unaffected by the base rate. The decision threshold is then optimized using out-of-fold training predictions to maximize balanced accuracy, which is the unweighted mean of the two class recalls.

\subsection{Results}

\input{Tables/prediction_results}

Table~\ref{tab:prediction} reports the results from two baseline methods and seven configurations of \name. The numbers are mean and standard deviation over five held-out folds. The main findings are as follows:

\paragraph{Gaze predicts perceived visual realism better than stimulus-only automated evaluation.}
The best accuracy model of \name, using all 24 saliency evaluation metrics, reaches 71.3\% accuracy, 0.713 balanced accuracy and 0.775 ROC-AUC---16 points of accuracy and 0.28 of AUC above the trivial baseline and 11 points of accuracy and 0.11 of AUC above the stimulus-only baseline.
This predictive ability originated from eye movement alone, as the model has no access to the video, the prompt or the generator's identity.
The bottom of Figure~\ref{fig:modeling} shows what the prediction results look like clip by clip, on a single partition in which every one of the 300 videos is scored exactly once by the fold that held it out.
The two rating classes separate but overlap substantially: the model is a useful ranking of videos by how real they looked, not a decision procedure that resolves individual borderline cases.
At the two ends of the axis the separation is clear. The running example \texttt{0036\_firefly}, a domino sequence whose viewers gave it a mean rating of 1.0, receives $p{=}0.05$; \texttt{0164\_sora2}, rated 4.5 on average, receives $p{=}0.95$.
Because that figure scores every clip once within one partition, while Table~\ref{tab:prediction} averages five independent splits, the two sets of numbers are close but not identical, and neither is a re-scoring of the other.
The model that combines saliency metrics features with VBench features demonstrates performance similar to the best model. It ranks second in both accuracy (68.7\%) and balanced accuracy (70.2\%), with the difference from the first model falling within one standard deviation. However, it outperforms the other models with the best ROC-AUC score of 0.805, which is 0.145 higher than the VBench features alone and 0.030 higher than the gaze features alone.
These results indicate that gaze data carries significant information that the automated stimulus features do not capture. Additionally, the stimulus features contribute a smaller yet consistent amount of information that enhances the gaze data.

\paragraph{Where gaze lands is informative, but how the eyes move by itself is not.}
Eye movement statistics alone has low performance (0.496 ROC-AUC, 0.503 accuracy) and are the only feature set that fails to beat the trivial baseline, even though the same four measures separated real from AI-generated clips significantly in Section~\ref{sec:study1}.
The two results are not contradictory, as those tests were paired within scene, so each scene acted as its own control for how much motion and clutter it contains, whereas the video-level model must compare across scenes, where raw fixation durations and saccade amplitudes are dominated by content.
In contrast, the saliency evaluation metrics do not have this problem, as each of them is already normalised against the clip in question---they compare each viewer to themselves, or to a prediction made for that same clip.
Adding the eye movement statistics features to the best saliency model does not help and in fact decreases 6 points of accuracy (0.650 vs.\ 0.713), as they contributing more variance than signal.

\paragraph{Self-consistency across the two passes is the strongest single family.}
Among the three saliency metrics blocks, each with eight features, the ranking is 1st vs.\ 2nd (0.751 AUC) $>$ 2nd vs.\ SalFoM (0.689) $>$ 1st vs.\ SalFoM (0.665).
The comparison a viewer makes against themselves is worth more than either comparison against a bottom-up model, and between those two the second pass is the more informative. This aligns with the ordering in the correlation analysis of Figure~\ref{fig:correlation_analysis} and in the group comparisons of Section~\ref{sec:study1}.
The three blocks are complementary rather than redundant. Together, they achieve an additional 2.6\% accuracy and a 0.024 increase in AUC compared to the best individual block. 
The findings show that the prediction of visual realism improves when a video is watched twice. This is the most significant feature in the data, as it cannot be computed from a single viewing alone. Each time the video is watched, it provides more information with the bottom-up saliency.

\subsection{How Many People Are Needed to Predict Visual Realism from Eye Tracking Data?}
\label{sec:viewer_model}

The \name models predict visual realism based on aggregated viewer ratings for a clip.
A further question is whether a person's gaze can predict their rating and how many viewers are needed to make a reliable prediction about visual realism.
We repeated the protocol using the best \name settings with saliency evaluation metrics (1st vs. 2nd vs. SalFoM).
We maintained all the steps the same in Figure~\ref{fig:modeling} but varied the number of viewers included in the modeling, ranging from $k=1$ to $k=8$.
Since each video in our dataset has average about 8 viewers,  the performance of the model with $k=8$ should align closely with the best results presented in Table~\ref{tab:prediction}.

\begin{figure}[!t]
\centering
\includegraphics[width=\linewidth]{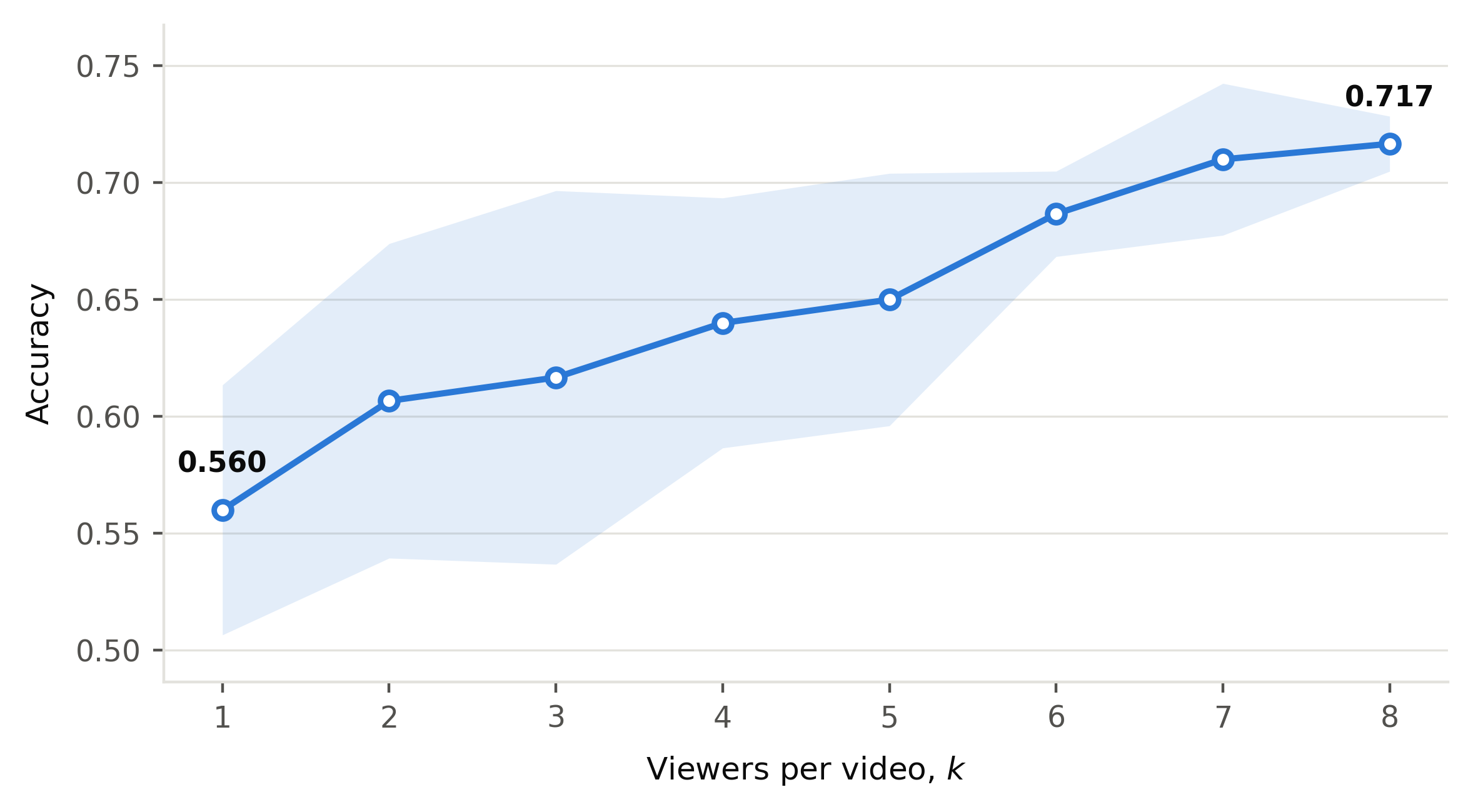}
\caption{Accuracy against the number of viewers per video for \name with saliency evaluation metrics (1st vs.\ 2nd vs.\ SalFoM). Accuracy rises monotonically with the number of viewers $k$ from $56.0\%$ ($k=1$) to $71.7\%$ ($k=8$). In our dataset, some clips have fewer viewers than $k$, so $k=7$ and $k=8$ average $6.88$ and
  $7.60$ viewers.}
\label{fig:prediction_k}
\end{figure}

The results of accuracy in relation to the number of viewers per video are illustrated in Figure~\ref{fig:prediction_k}. The data shows a monotonically increasing trend, reaching an accuracy of 71.7\%, which is close to our best results in Table~\ref{tab:prediction}. Additionally, the standard deviation decreases as the number of viewers increases.
For a single viewer, the accuracy is around 56\% with a ROC-AUC of 0.62. While this performance is above the trivial baseline, it is significantly lower than the model results achieved with multiple participants.
This difference is attributed to the noisy individual behavior and responses.
By averaging the inputs from multiple viewers, this noise is reduced a lot, which explains why the model with 8 viewers achieves an improvement of approximately 15\% in accuracy. 
Since the curve in the Figure~\ref{fig:prediction_k} hasn't converged, future research can continue scaling up the study to see the emerging trend.
The practical implication of this finding is that gaze-based evaluations should be considered at the audience level. Having more viewers per clip is promising for achieving reliable results. However, relying on only a few viewers may lead to unreliable outcomes.

%% file: Tables/prediction_results.tex
\begin{table*}[t]
  \centering
  \caption{Predicting perceived visual realism from gaze at video level. The input features are from video stimulus and eye tracking data. Scores are the mean $\pm$ standard deviation over 5 held-out folds. \#Features is the number of features. 
  Within each metric, the best score is colored in dark blue and the second is in light blue.}
  \label{tab:prediction}
  \small
  \begin{tabular}{lrccc}
    \toprule
    Method & \#Features & Accuracy & Bal.\ Accuracy & ROC-AUC \\
    \midrule
    Trivial baseline & -- & $0.550$ & $0.500$ & $0.500$ \\
    Stimulus-only baseline (VBench features) & 7 & $0.600 \pm 0.039$ & $0.614 \pm 0.044$ & $0.660 \pm 0.060$ \\
    \midrule
    Eye movement statistics & 8 & $0.503 \pm 0.069$ & $0.522 \pm 0.065$ & $0.496 \pm 0.096$ \\
    Saliency evaluation metrics (1st vs.\ SalFoM) & 8 & $0.617 \pm 0.073$ & $0.618 \pm 0.063$ & $0.665 \pm 0.076$ \\
    Saliency evaluation metrics (2nd vs.\ SalFoM) & 8 & $0.667 \pm 0.059$ & $0.665 \pm 0.057$ & $0.689 \pm 0.039$ \\
    Saliency evaluation metrics (1st vs.\ 2nd) & 8 & $0.687 \pm 0.034$ & $0.690 \pm 0.034$ & $0.751 \pm 0.046$ \\
    \textbf{Saliency evaluation metrics (1st vs.\ 2nd vs.\ SalFoM)} & 24 & \best{0.713 \pm 0.027} & \best{0.713 \pm 0.018} & \second{0.775 \pm 0.016} \\
    Saliency evaluation metrics (1st vs.\ 2nd vs.\ SalFoM) + statistics & 32 & $0.650 \pm 0.026$ & $0.660 \pm 0.027$ & $0.762 \pm 0.012$ \\
    Saliency evaluation metrics (1st vs.\ 2nd vs.\ SalFoM) + VBench & 31 & \second{0.687 \pm 0.036} & \second{0.702  \pm 0.034} & \best{0.805 \pm 0.021} \\
    \bottomrule
  \end{tabular}
\end{table*}

%% file: Sections/5-Conclusion.tex
\section{Conclusion and Discussion}

This paper investigated eye tracking as an implicit signal for evaluating visual realism of AI-generated videos.
We ran the two-pass Visual Turing Test with 90 participants (\textit{batch 1 \& 2}), and collected 4{,}286 valid eye-tracking trials on 438 videos from six advanced video generative models.
Our analysis shows that visual realism measurably changes how a video is watched: high visual realism content draws wider, more exploratory viewing patterns, while low visual realism content draws attention that settles early and returns to the same regions.
In the two-pass watching protocol, the first pass is usually dominated by bottom-up saliency, while the second pass is more influenced by top-down cognitive processing, as viewers search for anomalies.
Building on these findings, we trained the model \name that predict perceived visual realism from pure human eye-tracking data, reaching 71.3\% accuracy on videos, compared to the stimulus-only automated quality evaluation at 61.4\%.
The result establishes gaze as a complementary channel to explicit human feedback for evaluating video generation.
We publish this eye tracking dataset as a resource for further investigation, supporting the community in working toward more human-centered evaluation of generative video models.

\subsection{Limitations and Future Work}

We discuss three limitations of the current work and potential ways to improve them in future work.

First, our findings are limited by the video stimulus set and the viewing protocol we designed for our study. The video stimulus set only includes 5-second short videos featuring physical scenes without humans or narratives~\cite{motamed2025generative}. This represents a narrow aspect of video generation, as daily life and business scenarios often involve human presence. Additionally, the two-pass viewing protocol does not exactly reflect how videos are typically watched. While people may watch short videos multiple times, their primary focus is usually not on AI detection. This study serves as a starting point for investigating eye-tracking data as an evaluation signal. Future research should include a larger-scale crowdsourcing study in real-world settings to address these limitations.

Second, we used SalFoM~\cite{moradi2024salfom} for bottom-up saliency prediction, which was trained on natural videos. Its predictions may be systematically degraded when applied to AI-generated video clips, potentially affecting the results from the comparisons between human attention map and bottom-up saliency. This observed difference could be due to the saliency model’s failure to accurately assess AI-generated videos rather than reflecting true bottom-up saliency. To validate the video saliency prediction model, it is necessary to compare the agreement between SalFoM predictions and human attention maps from a free viewing task on AI-generated videos in the future.

Third, although \name has demonstrated that the spatial information derived from eye tracking data using eight saliency metrics is valuable for evaluation, this study does not consider the temporal aspects of human eye movements in its analysis and modeling. The sequence of where people look can provide insightful signals that may further improve the accuracy of visual realism prediction. For example, continuously focusing on one area for three seconds carries a different meaning than visiting that area for one second three times throughout the viewing period.
One future direction for \name is to integrate temporal or event-based eye tracking metrics. This could include scanpath metrics such as dynamic time warping, MultiMatch, ScanMatch, and Sequence Score~\cite{wang2026benchmarking}, as well as event-based metrics like revisit frequency and fixation transitions~\cite{shi2025chartist}. 
By incorporating these more comprehensive features from eye tracking data, a future model may achieve improved accuracy in predicting visual realism.


%% file: Sections/appendix.tex
\newpage
\onecolumn
\appendix

\section{Full Group Comparison Statistics}
\label{sec:appendix_stats}

\begin{figure*}[!h]
\centering
\includegraphics[width=\linewidth]{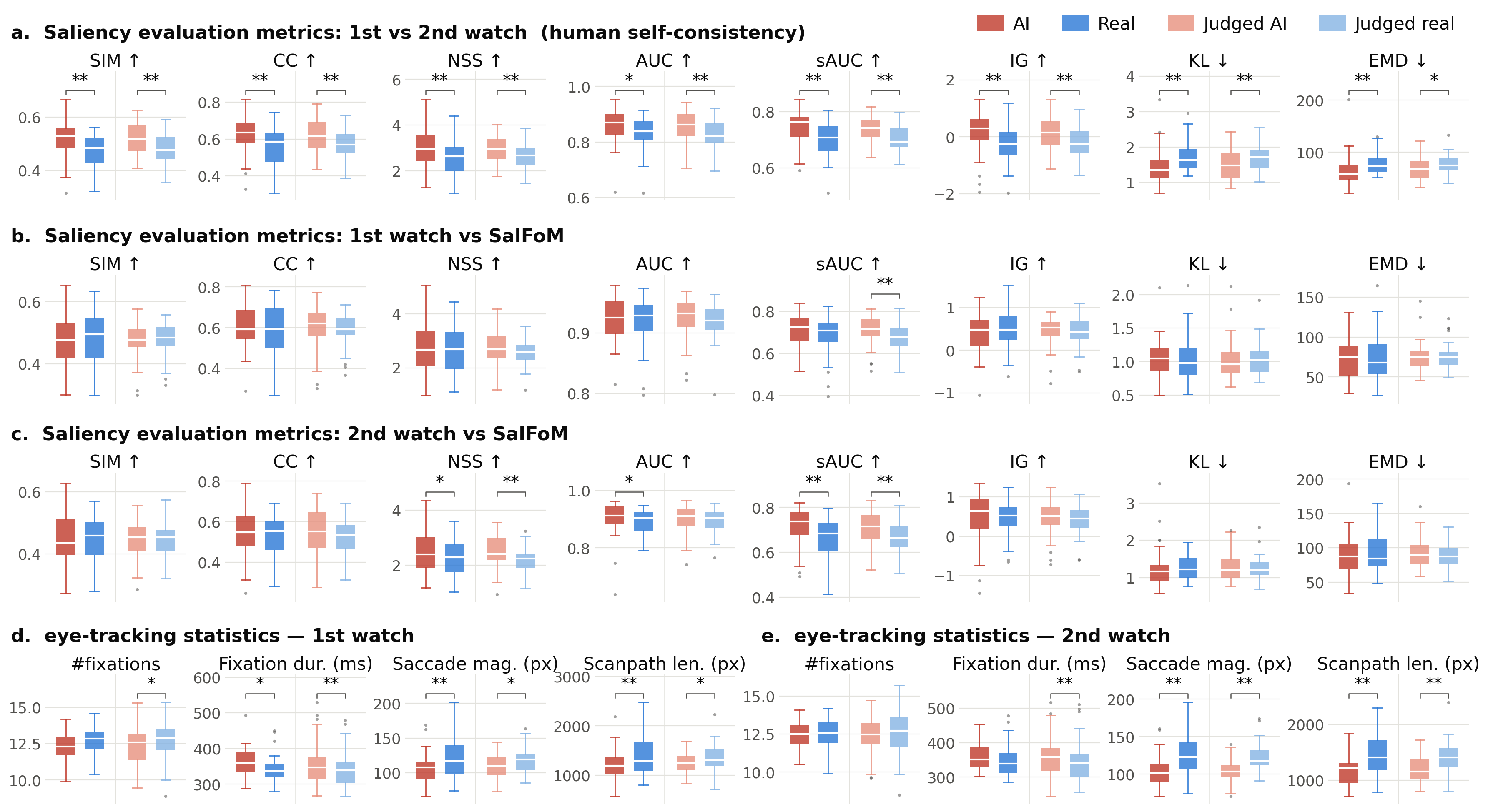}
\caption{Group comparisons for the two-pass Visual Turing Test. (a--c) report the eight saliency evaluation metrics for the three attention-map comparisons; (d--e) report the four eye movement statistics for each pass. 
Each panel carries two comparisons: AI vs.\ real clips and judged AI vs.\ judged real. Brackets mark Wilcoxon signed-rank tests: $*\,p<.05$, $**\,p<.01$. }
\label{full_analysis_extended}
\end{figure*}

The full data extended Figure~\ref{full_analysis} is plotted in Figure~\ref{full_analysis_extended}.
The findings drawn from it are discussed in Section~\ref{sec:study1}.

